\documentclass[graybox]{svmult}

\usepackage{helvet}         
\usepackage{courier}        
\usepackage{type1cm}        
\usepackage{makeidx}         
\usepackage{graphicx}        
\usepackage{multicol}        
\usepackage[bottom]{footmisc}

\usepackage{xcolor}
\usepackage{graphicx} 
\usepackage{booktabs} 
\usepackage{subfig}
\usepackage[english]{babel}
\usepackage{amsfonts}
\usepackage{amssymb}
\usepackage{amstext}
\usepackage{amsmath}
\usepackage{amsmath,bm}

\usepackage{multicol}
\usepackage{mwe,tikz}\usepackage[percent]{overpic}

\newcommand\mba{\mathbf{a}}
\newcommand\mbb{\mathbf{b}}
\newcommand\bee{\begin{equation*}}
\newcommand\eee{\end{equation*}}
\def\be{\begin{equation}}
\def\ee{\end{equation}}
\def\ba{\begin{eqnarray}}
\def\ea{\end{eqnarray}}

\def\nn{\nonumber}
\newcommand\dphi{\dot{\phi}}
\newcommand\eb{b e^{-\alpha\phi}}
\def\a{\mathbf{a}}

\def\r{\bar{\rho}}

\begin{document}

\title*{Cosmological solutions from models with unified dark energy and dark matter and with inflaton field}
\author{Denitsa Staicova and Michail Stoilov}
\titlerunning{Cosmological solutions of two-measures models}
\institute{Denitsa Staicova \at Institute for Nuclear Research and Nuclear Energy \\Bulgarian Academy of Sciences, Sofia 1784, Tsarigradsko shosse 72, Bulgaria, \email{dstaicova@inrne.bas.bg}
\and Michail Stoilov \at Institute for Nuclear Research and Nuclear Energy \\Bulgarian Academy of Sciences, Sofia 1784, Tsarigradsko shosse 72, Bulgaria \email{mstoilov@inrne.bas.bg}}
\maketitle

\abstract{Recently, few cosmological models with additional non-Riemannian volume form(s) have been proposed.
In this article we use  Supernovae type Ia experimental data to test one of these models which provides a unified description of both dark energy via dynamically generated cosmological 
constant and dark matter as a "dust" fluid due to a hidden nonlinear Noether symmetry. It turns out that the model allows various scenarios of the future Universe evolution and in the same time 
perfectly fits contemporary observational data.
Further, we investigate the influence of an additional inflaton field with a step like potential. With its help we can reproduce the Universe inflation epoch, matter dominated epoch and present 
accelerating expansion in a seamless way. Interesting feature is that inflaton undergoes a finite change during its evolution. It can be speculated that the inflaton asymptotic value is connected to 
the vacuum expectation value of the Higgs field.}

\section{The two-measures model}
The application of the two-measures model \cite{ref01, ref01_1, ref01_2, ref01_3} to cosmology has been pioneered in series of articles by Guendelman, Nissimov and Pacheva 
(\cite{1407.6281, 1408.5344,1505.07680,1507.08878,1508.02008,1511.07071,1603.06231,1609.06915}). In those articles, it has been described a model which is able to describe both dark matter and dark 
energy in the Universe and also early inflation. This is achieved by the introduction of two scalar fields -- a darkon and an inflaton -- in a scalar Lagrangian coupled both to the standard Riemannian 
volume-form (the square root of the metric determinant) and to another non-Riemannian volume form (given in terms of auxiliary maximal-rank antisymmetric tensor gauge field). The effect of the 
additional measure in the theory is felt only through the ratio of the two measures -- a constraint determined by an algebraic equation. The equations of motion of such a theory generate dynamically a 
cosmological constant and a dark matter dust fluid term and also inflation-inducing terms. 

In this article, we will discuss our recent numerical investigations of this model in the case of both darkon-only universe and of darkon-inflaton universe. In the first case, we were able to 
successfully fit the model with the Supernova Type 1 data and to limit its parameter space to observationally acceptable values. We also showed that in this case it is possible for the Universe to 
undergo a phase transition.  In the second case, we were able to reproduce the stages of the Universe expansion -- early inflation, matter domination and late inflation under certain choices for the 
parameters. We also observed some novel features, like the matter-dominated early epoch and a non-zero scalar field in the late Universe. In both cases, we have numerically confirmed that the 
two-measures model can be a viable cosmological model.

\section{The darkon model in FLRW metric}
The action of the two-measures darkon model in the f(R) gravity (Guendelman, Nissimov and Pacheva (\cite{1408.5344},\cite{1508.02008})) has the following form:
$$S_{darkon}=\int d^4x\sqrt{-g}(R(g,\Gamma)-\alpha R^2(g,\Gamma))+\int{d^4x(\sqrt{-g}+\Phi(C))L(u,X)}$$
where $\Phi(C)=\frac{1}{3\!}\epsilon^{\mu\nu\kappa\lambda}\partial_\mu C_{\nu\kappa\lambda}$ is the non-Riemannian measure and 
$L(u,X)=-\frac{1}{2}g^{\mu\nu}\partial_\mu u\partial_\nu u-V(u)$ is the matter Lagrangian of the darkon scalar field $u$.

If one applies the equations of motion obtained from this action to the Friedman--Lema\^{i}tre--Robertson--Walker metric with $k=0$:
\be
ds^2 = - dt^2 + a(t)\left[dr^2 + r^2 \left( d\theta^2 + \sin^2\theta d\varphi^2\right)\right] .
\label{FLRW-metric}
\ee
one obtains from the Friedman equations ($G_{00}=T_{00}$) the following relations for the energy density:
\begin{align}
&\rho= \frac{1}{8\alpha}\dot{u}^2 +\frac{3}{4}\frac{p_u}{a(t)^3}\dot{u}-\frac{1}{4\alpha} \label{rho} \\
&p_u = a(t)^3\left[ -\frac{1}{2 \alpha}\dot{u}+(\frac{1}{4\alpha}-2M_0) \dot{u}^3\right] \label{em1}
\end{align}
where $p_u=const$.

Following our work in \cite{num_dark}, we rewrite the last cubic equation for $\dot{u}$ (Eq. (\ref{em1})), as 
$y^3 +3 \mba y + 2 \mbb = 0$ with $\mba=-\frac{2}{3 - 24 \alpha M_0}$ and $\mbb= -\frac{2\alpha p_u}{a(t)^3(1-8\alpha M_0)}$, $y=\dot{u}$.
    
The solutions are:    
\ba 
y_1&=&-\frac{\mba}{(-\mbb+\sqrt{\mba^3+\mbb^2})^{1/3}}+ 
(-\mbb+\sqrt{\mba^3+\mbb^2})^{1/3}
\label{y_1}\nonumber\\
y_2&=&\frac{\mba}{(\mbb-\sqrt{\mba^3+\mbb^2})^{1/3}}- 
(\mbb-\sqrt{\mba^3+\mbb^2})^{1/3}
\label{y_2}\nonumber\\
y_3&=&\frac{y_2-i\sqrt{3}y_1}{2}
\label{y_3}\nonumber
\ea

Since no real smooth solution exists in the whole [$\mba, \mbb$] plane, we define the following piecewise functions, {\bf real} in the whole plane [$\mba, \mbb$]:

\begin{multicols}{2}\noindent
\begin{equation*}
y_b=\begin{cases}
y_1 \;\;\mathrm{for}\;\; (a,b)\in\{a\geq 0\}\cup\{a < 0 \cap b<0\}\\
y_2\;\; \mathrm{for}\;\; (a,b)\in \{a < 0 \cap b>0\}\label{yp}
\end{cases}
\end{equation*}
\begin{equation*}
\; \; \; \; \; \; y_s =\begin{cases}
\; y_1\;\; \mathrm{for}\;\; b>0\\
\; y_2\;\; \mathrm{for}\;\; b<0.
\end{cases}
\end{equation*}
\end{multicols}

We obtain the final form of the Friedman equation after rescaling time by $2|\alpha|/3 =1$ and absorbing $\alpha$ into Hubble constant ($\bar{\rho}=4|\alpha|\rho$): 

\be 
\left(\frac{\dot{a}(t)}{a(t)}\right)^2=\bar{\rho}=\left(\frac{1}{2} y^2 + \frac{\mbb}{\mba} y -1\right)
\label{freedman}
\ee
The asymptotics, corresponding to the dark energy term in the late universe, is:
\begin{equation*}
\r \xrightarrow[a(t)\rightarrow\infty]{}
\begin{cases}
 1 \;\; \mathrm{for}\;\;\a>0\nn\\
-\frac{3}{2}\a-1 \;\; \mathrm{for}\;\;\a<0\label{asym}
\end{cases}
\end{equation*}

We use as independent real solutions $y_b$ (our basic solution) and  $y_s$ and integrate numerically Eq. (\ref{freedman}) to find the evolution of 
the universe.

{\bf Phase transition:} From the numerical integration we have seen that it is possible to obtain both Universes with or without phase transition. The possibility for such transition comes from the 
fact that in the $[\mba,\mbb]$ plane exist sectors where two of our solutions have positive energy density $\rho$. 

Explicitly, let's denote $\bar{\bar{\rho}}$ the density corresponding  to solution $y_s$ ($\bar{\rho}$ corresponds to $y_b$). At the moment $t_1=0$, it will be negative, i.e. 
$\bar{\bar{\rho}}(t_1)<0$. For certain moment $t_p$, however, it will change sign: $\bar{\bar{\rho}}(a(t_p))=0$. Therefore, for any moment $t>t_p$ we have two "states" of the Universe $\r$ and 
$\bar{\bar{\rho}}$ such as:
\be  
0\leq\bar{\bar{\rho}}<\r\;\; \mathrm{for}\;\; t\geq t_p. 
\ee
This opens the possibility the Universe to undergo {\em "phase transition"} or {\em "quenching"} to the lower state.

\begin{figure}%
    \centering
    \subfloat{{\includegraphics[width=7cm]{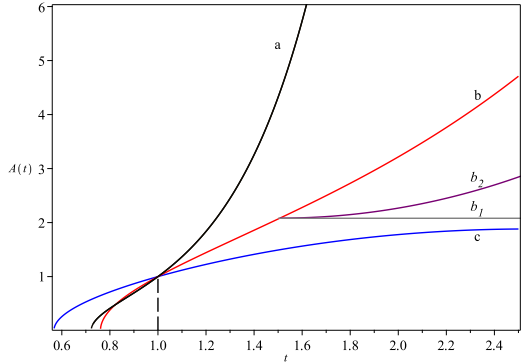} }}%
\caption{Graphics of the $a(t)$ evolution for: $\mba= - 5.987, \mbb=\frac{-2.932}{a^3}$ (a) $\mba=-1, \mbb=-\frac{2}{a^3}$ (b),
 $t_p=1.5074,a_s(t_p)=2.0825$ ($b_2$), $\mba=-.5, \mbb=-\frac{0.5}{a^3}$ (c) }
 \label{evol_dark}
    \end{figure}

The moment of  the phase transition is crucial for the further evolution, since if $t = t_p$ the evolution stops ($\bar{\bar{\rho}}=0$), if $t = t_p +\delta t$,  we observe phase transition 
of the first kind. An illustration of this process can be seen on Fig. \ref{evol_dark}, where the transition happens between lines $b, b_1$ and $b_2$. 

{\bf The Supernova Fit:} Using the freely available data of Supernovae Type 1a \cite{sn}, we were able to fit the distance modulus $d_m$\footnote{$d_m=5 \log_{10}\left(\frac{d}{10}\right)$, where 
$d$ is the distance in parsecs.} as a function of z using using energy density from the two-measures model:
 \ba
d_m &=& 5 \log_{10}\left((1+z)\int_0^z dx \frac{a(x)}{\dot{a}(x)}\right)=\nonumber\\
&&\mathrm{const}+5 \log_{10}\left((1+z)\int_0^z dx \frac{1}{\sqrt{\rho(x)}}\right)
\label{dm_b}
\ea

Using a symplectic fit, we were able to prove that our model is able to reproduce the observational data. The details on the fit can be seen in \cite{num_dark}, here we will emphasize only that the 
precision of the fit ($\chi^2 \sim 562 \;\;\mathrm{for} \;\; \a<-2/3, \chi^2 \sim 578 \;\;\mathrm{for} \;\; \a>1.$) is similar to the one of the standard model ($\chi^2=562$).

The best fit of SN data using the proposed model is not unique. We find one parametric family of solutions producing the same $d_m(z)$ function.
An approximate formula for the dependency $\mbb(\mba$) can be obtained using the LeastSquares algorithm in Maple and it gives: $$b_\pm=\sum_0^{4} \pm c_i a^i + O(a^5), \text{with coefficients}$$
\footnotesize 
$c_i = [0.337906, 0.376679, -0.0251697, 0.00148545, 0.112727 10^{-3}]$
\normalsize

On Fig. \ref{evol_dark}, curve (d) represents one such evolution with parameters $\mba$ and $\mbb$ corresponding to the observational data.

\section{The two-measures theory -- including the inflaton}
In order to produce inflation in the model, one needs to include a new scalar field -- the inflaton $\phi$. Following Guendelman, Nissimov and Pacheva \cite{1609.06915,1408.5344} (where in $S_{darkon} 
\; \alpha=0)$), the action, featuring two non-Riemannian measures $\Phi_1(A)$ and $\Phi_2(B)$, becomes:
\footnotesize
$$S=S_{darkon}+\int d^4x \Phi_1(A)(R+L^{(1)})+\int d^4x\Phi_2(B)\left(L^{(2)}+\frac{\Phi(H)}{\sqrt{-g}}\right)$$
\normalsize
where: 
\footnotesize
\begin{align}
& L^{(1)}=-\frac{1}{2}g^{\mu\nu}\partial_\mu\phi\partial_\nu \phi- V(\phi),\; V(\phi)=f_1 e^{-\alpha \phi} \\
&L^{(2)}=-\frac{b}{2}e^{-\alpha\phi} g^{\mu\nu}\partial_\mu\phi\partial_\nu \phi + U(\phi),\; U(\phi)=f_2 e^{-2\alpha \phi}\\
\end{align}
\normalsize
From the equations of motion we have:
\footnotesize
\begin{align*}
&p=-2M_0=const, \frac{\Phi_2(B)}{\sqrt{-g}}=\chi_2=const\\
&R+L^{(1)}=-M_1=const, L^{(2)}+\frac{\Phi(H)}{\sqrt{-g}}=-M_2=const\\
&U_{eff}(\phi)=\frac{(V_1(\phi)+M_1)^2}{4\chi_2(U(\phi)+M_2)} \text{with } U_{-}=\frac{f_1^2}{4\chi_2 f_2}, U_{+}=\frac{M_1^2}{4\chi_2 M_2}
\end{align*}
\normalsize
An important condition following from the requirement that the vacuum energy density of the early Universe $U_-$ should be much higher than that of the late Universe $U_+$ gives:
\begin{equation}
 \frac{f_1^2}{f_2}>>\frac{M_1^2}{M_2}
 \label{en_cond}
\end{equation}
This ensures that the effective potential has the form of two infinite plateaus connected with a steep slope. 

Additionally, one can postulate:
$$|M_1|\sim M_{EW}^4,M_2\sim M_{Pl}^4, f_1\sim f_2 \sim 10^{-8}M_{Pl}^4,$$
 so that one can connect the theory with the electroweak and the Planck scales.

The system of equations that need to be solved numerically in order to obtain the evolution of the Universe is the following:
\footnotesize
\begin{align}
&v^3+3av+2b=0 \text{ for }\\
&a_{} = \frac{-1}{3}\frac{V(\phi)+M_1-\frac{1}{2}\chi_2 b e^{-\alpha\phi}\dot{\phi}^2}{\chi_2(U(\phi)+M_2)-2M_0}, b_{} = \frac{-p_u}{2a(t)^3(\chi_2(U(\phi)+M_2)-2M_0)}\label{sys1}\\
&\dot{a}(t)=\sqrt{\frac{\rho}{6}}a(t), \; \rho_=\frac{1}{2}\dphi^2 (1+\frac{3}{4}\chi_2 b e^{-\alpha\phi} v^2)+\frac{v^2}{4} (V+M_1)+\frac{3 p_u v}{4a(t)^3} \label{sys2}\\
&\ddot{a}(t)=-\frac{1}{12}(\rho+3p)a(t), \; p_{}=\frac{1}{2}\dphi^2 (1+\frac{1}{4}\chi_2\eb v^2)-\frac{v^2}{4}(V+M_1)+\frac{p_u v}{4 a(t)^3}\label{pr}\\
&\frac{d}{dt}\left( a(t)^3\dphi(1+\frac{\chi_2}{2}\eb v^2) \right)+a(t)^3 (\alpha\frac{\dphi^2}{4}\chi_2\eb v^2+\frac{1}{2}V_\phi v^2-\chi_2 U_\phi\frac{v^4}{4})=0\label{sys3}
\end{align}
\normalsize
Here Eq. (\ref{pr}) is optional and it offers an independent way to evaluate $\ddot{a}(t)$. This differential system is of first order with respect to $a(t)$ and of second order with 
respect to $\phi(t)$. Once again, we first solve the cubic equation by choosing a base solution and then we use it, to integrate the differential system with the 
implemented in Maple Fehlberg fourth-fifth order Runge-Kutta method with degree four interpolant.

\begin{figure}[!ht]
       \centering
       \subfloat{{\includegraphics[width=7cm]{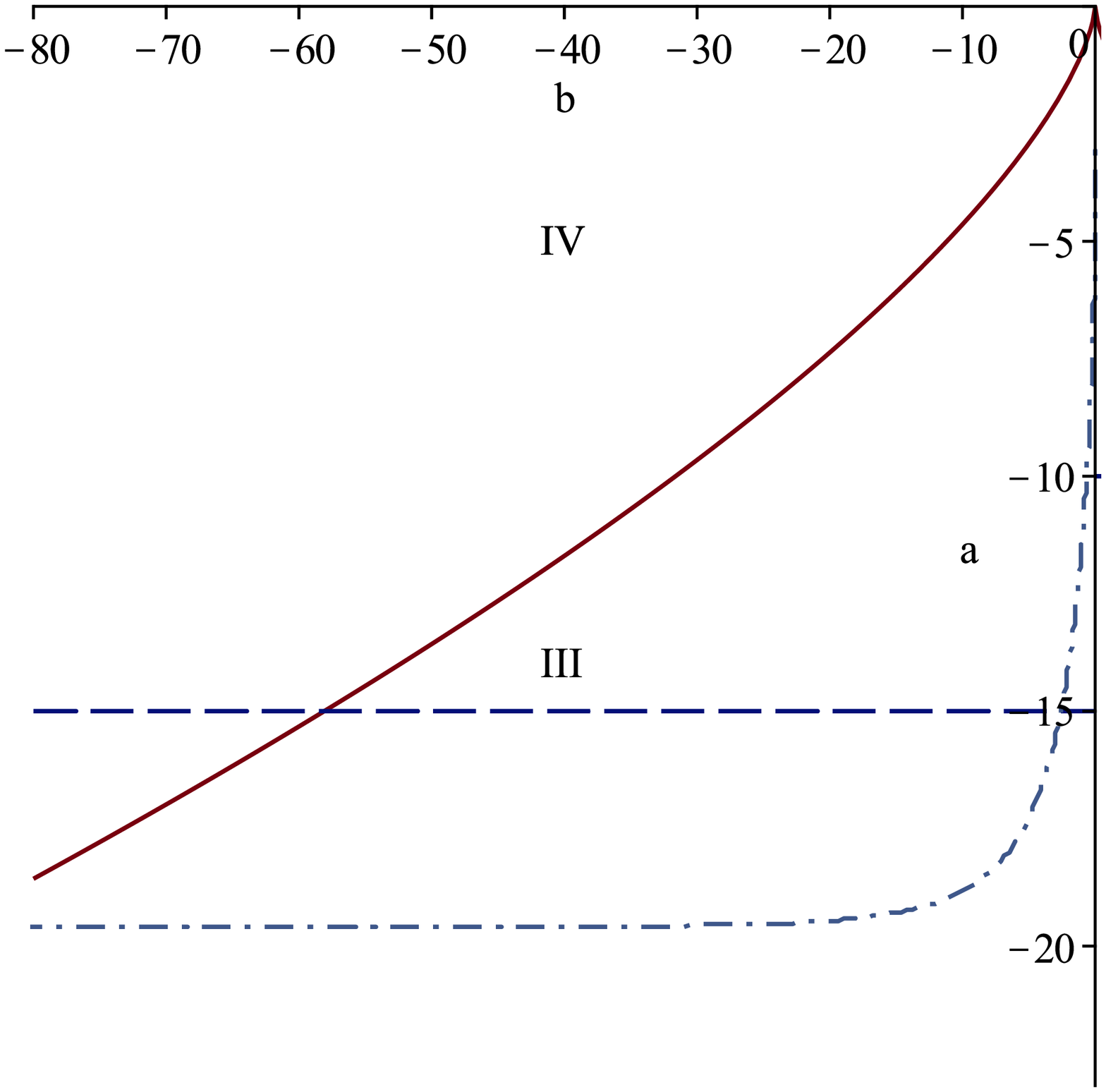} }}%
       \caption{The evolution of the solutions in the $[\mba, \mbb]$ plane. The dashed line correspond to the darkon case, the dash-dot line -- to the inflation case, the solid lines -- the lines of 
validity of the solutions. }
\label{ab}
     \end{figure}

{\bf Study the $[\mba, \mbb]$ plane} Unlike the previous case, where $\mba$ was a constant, here it depends on $\phi, \dot{\phi}$. 
Because of this, the trajectories in the $[\mba, \mbb]$ plane which the Universe will describe in its evolution won't be straight lines like in the darkon case, but curves. For example, on Fig. 
\ref{ab} we have plotted the trajectory for one set of parameters (dots). It starts at $b\to -\infty$ and ends at $b\to 0$. In its evolution, it crosses the $a^3+b^2=0$ line (solid 
line). On the plot, one can see also the trajectories for the darkon case plotted with dashed lines. In order to work in the sector III, where both solutions $y_b$ and $y_s$ are valid we have chosen 
the parameters in such a way that $b=\frac{-p_u}{2a(t)^3(\chi_2(U+M_2)-2M_0)}<0$. 

{\bf Numerical integration and equation of state} For the numerical integration, we impose as initial conditions: $$a(0)=10^{-12}, \phi(0)=\phi_0, \dot{\phi}(0)=0.$$
Additionally, we impose the gauge condition: $a(1)=1$.

This problem has 12 parameters of the system. Our numerical experiments show that the system is extremely sensitive to them and small changes can lead to either eternally exponentially expanding 
Universe or collapsing without inflation Universe. In order for an evolution to reproduce the known past of the Universe the second derivative of the scale factor 
$a''(t)$ has to change sign at least 2 times: to be positive during early inflation $a_{i}''(t)>0$, to be negative during matter domination period ($a_{MD}''(t)<0$, and to be positive during the late 
(current) expansion $a_{LE}''(t)>0$. 

\begin{figure}
 \subfloat[]{{\includegraphics[width=5.63cm]{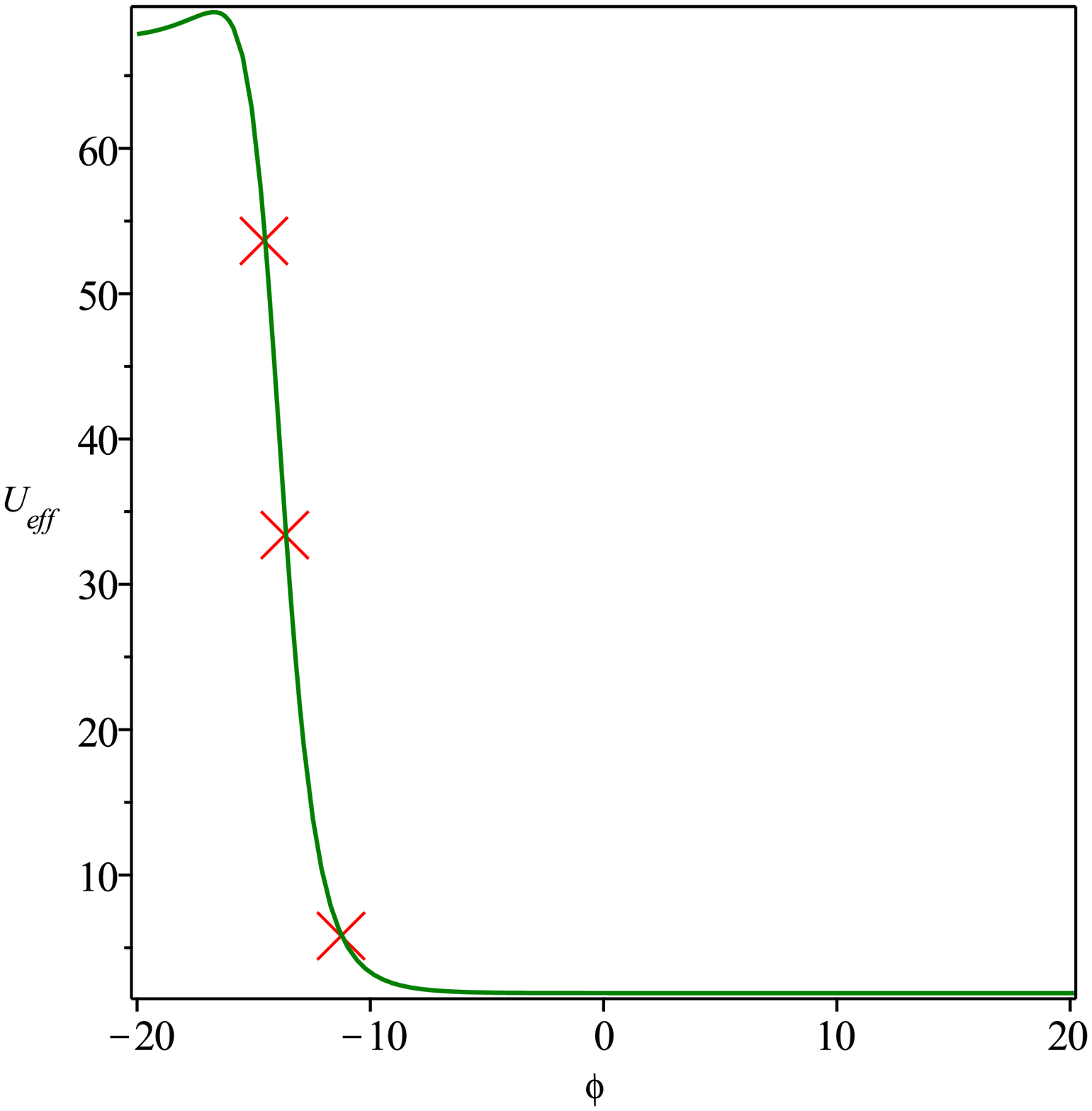} }}%
 \quad
  \subfloat[]{{\includegraphics[width=5.63cm]{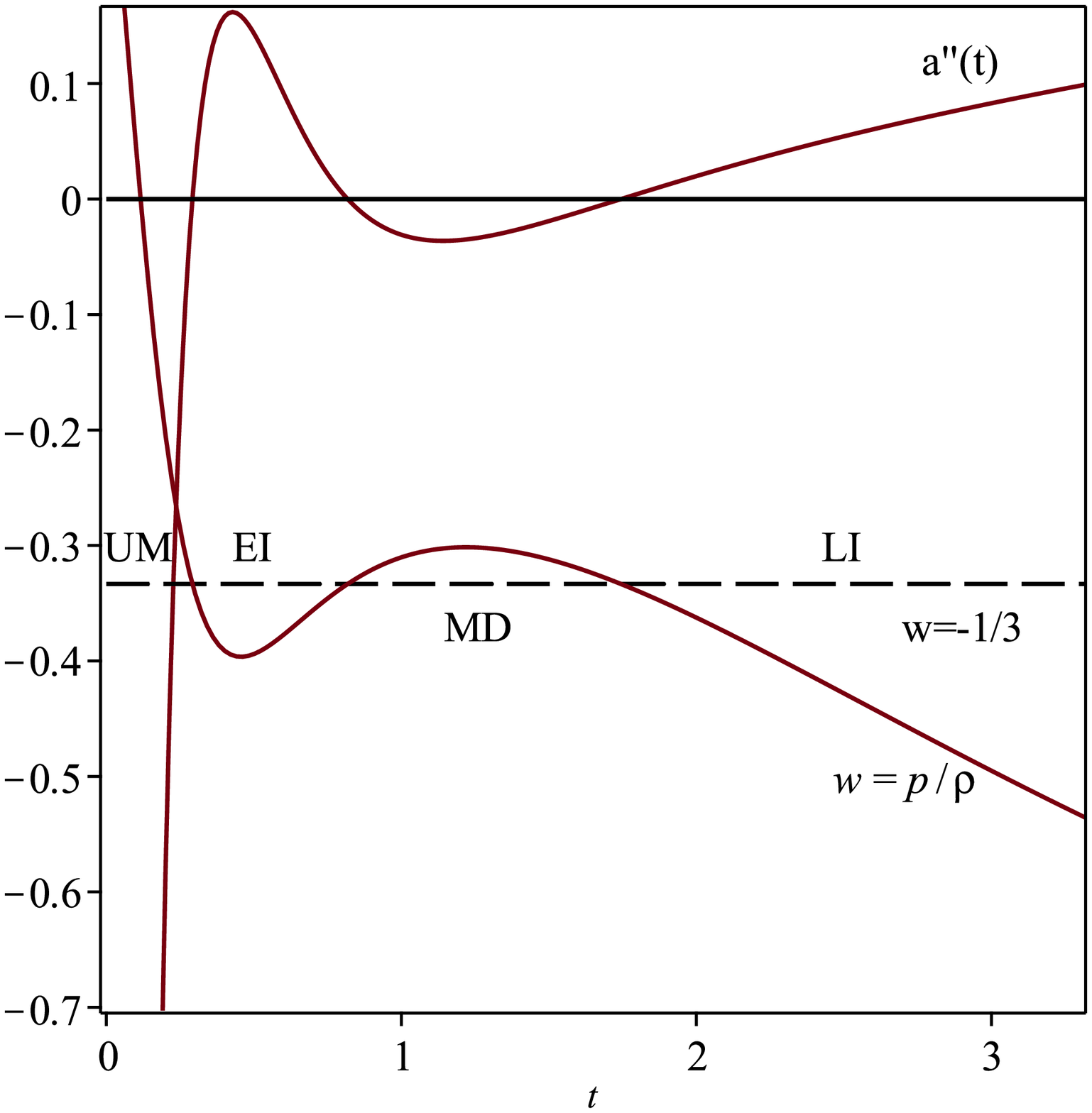} }}%
\caption{ a) The effective potential, where the crosses signify the moments: $t=0$, $t=1$ and $t=4$. b) Plot of $\ddot{a}(t)$ and the equation of state $w=p/\rho$. One can see the different 
epochs -- ultra-relativistic matter domination (UM), the early inflation (EI), the matter domination (MD) and the late inflation (LI).  }
\label{evol}
\end{figure}

Our numerical investigations show that such ``physical`` cases are indeed possible, for example Fig. \ref{evol}, but require careful fine-tuning of the parameters .\footnote{The plots are for 
parameters $M_0 = -0.01, M_1 = 0.1, M_2 = 4, \alpha = .7, b_0 = 1\times 10^{-5}, p_u = .15, \chi_2 = 3.3\times 10^{-4}, f_1 = 3\times 10^{-5}, f_2 = 1\times 10^{-8}$, integrated for $t=0..4$ }. One 
can see the different epochs by plotting the equation of state $w=p/\rho$ (see Fig. \ref{evol} b). The times in which they kick in correspond to the change of sign of $\ddot{a}(t)$. 

A notable result from our work is that it is not numerically possible to start from the left plateau and to obtain a ''physical`` evolution. Instead, the evolution explodes to eternal inflation. It 
is not possible also to finish on the right plateau, because the evolution of the scalar field stops before reaching it (there is a friction term). This illustrated on Fig. \ref{evol} (a), 
where one can see the effective potential in this case.

\begin{figure}
   \subfloat[]{{\includegraphics[width=5.63cm]{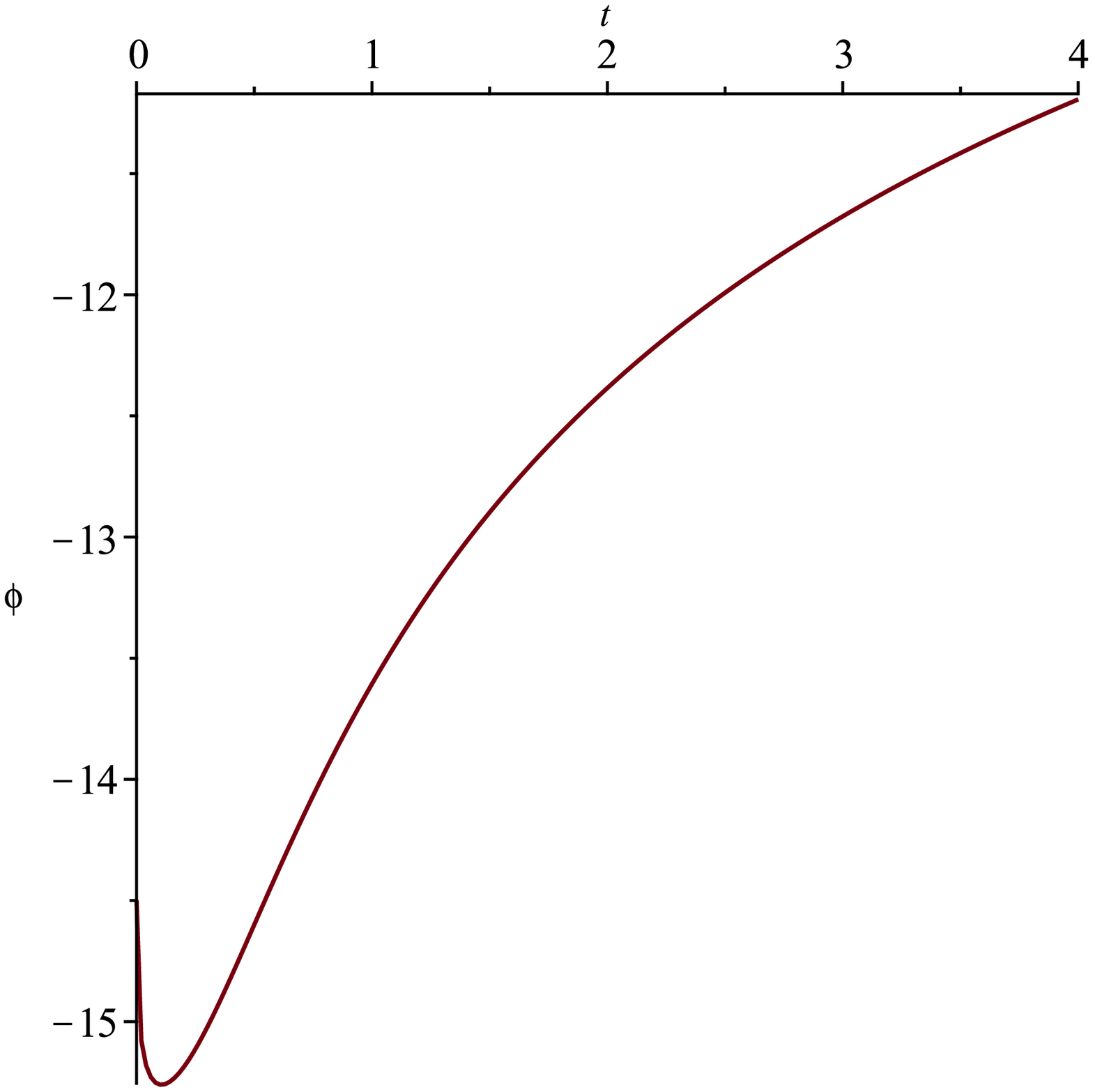}}  }
 \quad
  \subfloat[]{{\includegraphics[width=5.63cm]{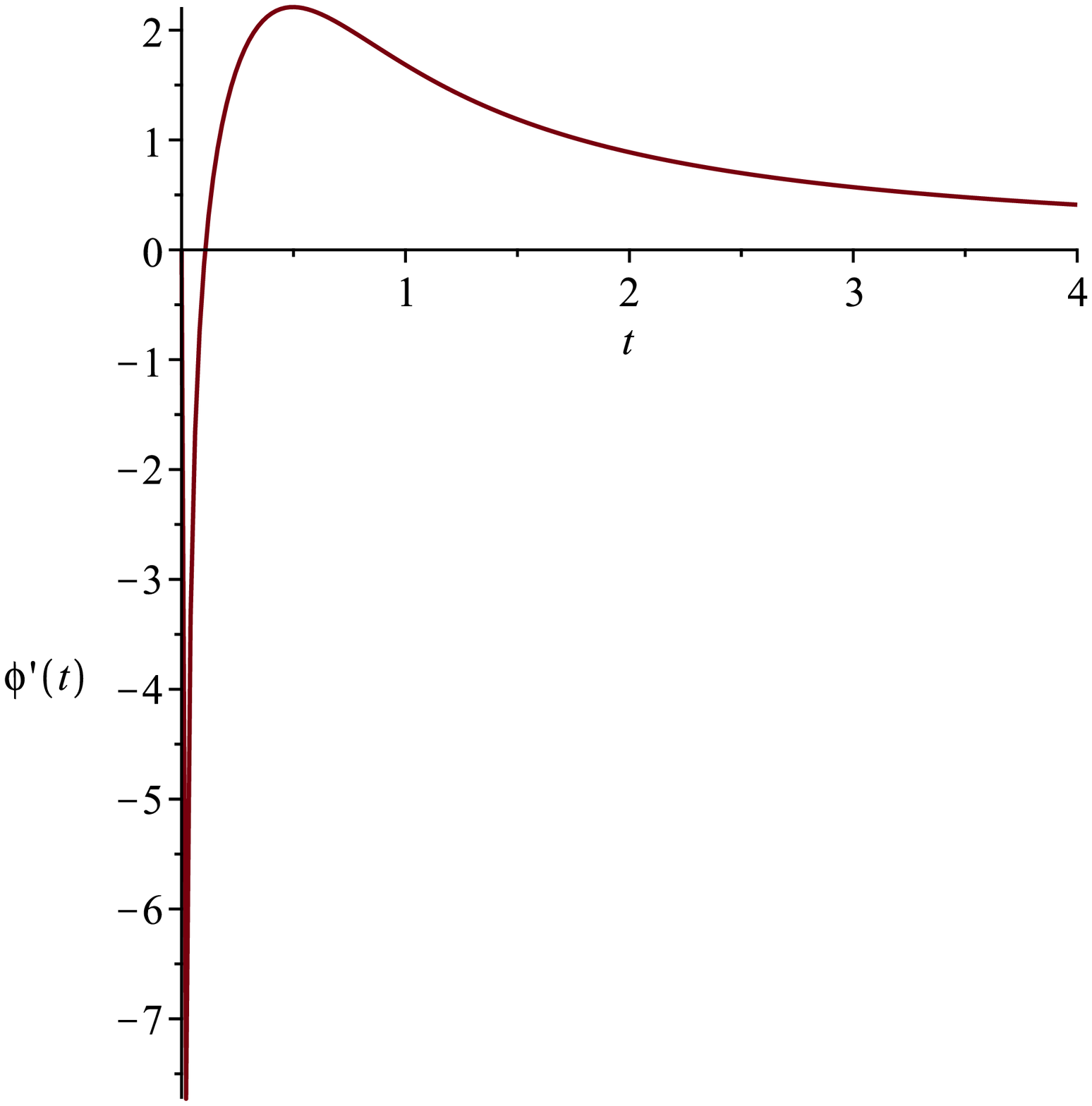} }}%
\caption{ a) The inflaton scalar field $\phi(t)$ b) Its derivative $\dot{\phi}(t)$ }
\label{evol_phi}
\end{figure}

A very important feature of the model is that it starts with a pre-inflation mater domination epoch with exponentially high energy density, which quickly cools to enter in the early inflation stage 
(see Fig. \ref{evol} b) ). Another important feature is the fact that the scalar field does not reach zero in the late Universe as expected by the theory. This is also due to the friction term which 
stops its evolution ($\dot{\phi}=0$) before it can reach zero (Fig. \ref{evol_phi}). 

In our numerical simulations we have discovered some interesting features of the model, which is not in accordance with the asymptotic found in \cite{1408.5344}. One needs to keep in mind, however, 
that due to the numerical complexity of the problem and its big parameter-space, we have not yet reached the theoretically predicted values of some of the parameters discussed in there.
While the main requirement of the model Eq. (\ref{en_cond}) is satisfied, the values of the other parameters, which can be found in the table \ref{tbl}, should be extended to a more physically 
realistic domain. 

\begin{center}
\begin{tabular}{|l|l|l|}
\hline
Parameter & Theory & Numerics \\
\hline
$M_1$ & $\sim M_{EW}^4=4.10^{-60}$ & {$1/15=6.67\times 10^{-2}$}\\
$M_2$ & $\sim M_{Pl}^4=4$ & $4$  \\
$f_1$ & $\sim  10 ^{-8}$ & {$2\times 10^{-5}$}   \\
$f_2$ & $\sim  10 ^{-8}$ & $10^{-8}$  \\
$M_0$ & $\Lambda^{Pl}\sim 10^{-122}$ & {$\Lambda=1.156\times 10^{-5}$} \\
$\alpha$ & $10^{-20}-0.2$ & 	{$0.64$} \\
\hline
\end{tabular}
\label{tbl}
\end{center}

Another way to define ``physicality'' of the problem, is the problem of the time scales. The observationally expected values for the time at which $\ddot{a}(t)$ changes its sign comes from the time 
when different epochs start. Theoretically, matter domination is considered to start at $a_{MD}(t)\sim 3\times 10^{-4}$ and the accelerated expansion -- at $a_{AE}(t)\gtrsim 0.6$. Our current best 
result is $a_{MD}=0.2, a_{AE}=1.2$. It is yet to be seen whether the observational values can be reached trough fine-tuning of the parameters. Because of the complexity of the problem, 
this fine-tuning needs to be done step by step and cannot be automatized for the moment. 

Finally, due to the extreme predicted ratio $U_+/U_-\sim 10^{120}$, reaching the theoretically predicted values of the parameters may be computationally impossible, due to possible increase in the 
 required precision for the numerical integration of the system. A fuller investigation of the parameter space of the problem will be presented in future works.

\section{Conclusions}
In our numerical work on the application of the two-measures model to cosmology we have confirmed that this model can be considered as an alternative of the standard model of dark matter and dark 
energy. Through numerical integration of the Friedman equations in the K-essence theory in the darkon and the inflaton case, along with detailed study of the plane $[\mba, \mbb]$, we have 
obtained interesting numerical results. 

It was shown that in the darkon model we can obtain both a Universe with and without phase transition and those models were fit to the data of Supernovae Type 
1a. In the case of inflaton model, we have performed first steps in the study of the the parameter space of the model and we have found solutions for which one can obtain the two 
inflationary epochs and one matter dominated epoch. It was shown that the inflation experiences friction, due to which inflation stops before reaching the $U_	{+}$ part of the potential. This 
was unexpected result which is to be further investigated, because it also means that there should be a non-zero scalar field surviving to the modern epoch. 

\section*{Acknowledgments}

It is a pleasure to thank E. Nissimov and S. Pacheva for the discussions. 

The work is supported by BAS contract DFNP -- 49/21.04.2016, by Bulgarian NSF grant DN-18/1/10.12.2017 and by Bulgarian NSF Grant 08-17.


\begin{thebibliography}{99}
\bibitem{ref01} E.I. Guendelman, Mod. Phys. Lett. A {\bf 14} 1043-1052 (1999), arXiv:gr-qc/9901017;

\bibitem{ref01_1} E. Guendelman  and  A.  Kaganovich, Phys.  Rev. D {\bf 60} 065004 (1999), arXiv:gr- qc/9905029; 

\bibitem{ref01_2} E. Guendelman and O. Katz,  Class. Quantum Grav. {\bf 20} 1715-1728 (2003), arXiv:gr-qc/0211095

\bibitem{ref01_3} E.I. Guendelman and P. Labrana, Int. J. Mod. Phys. D22 (2013) 1330018, arxiv:1303.7267 [astro-ph.CO];
E.I. Guendelman, D. Singleton and N. Yongram, JCAP 1211 (2012) 044, arxiv:1205.1056 [gr-qc];
E.I. Guendelman, H.Nishino and S. Rajpoot, Phys. Lett. 732B (2014) 156, arXiv:1403.4199 [hep-th].

\bibitem{1407.6281} E.I. Guendelman, E. Nissimov and S. Pacheva, {\it Unification of Inflation and Dark Energy from Spontaneous Breaking of Scale Invariance}, "Eight Mathematical Physics Meeting'', 
pp.93-103, B. Dragovic and I. Salom (eds.), Belgrade Inst. Phys. Press (2015) {\bf 10}, arXiv:1407.6281 [hep-th]
%
\bibitem{1408.5344}E. Guendelman, R. Herrera, P. Labrana, E. Nissimov, S. Pacheva, {\it Emergent Cosmology, Inflation and Dark Energy}, General Relativity and Gravitation {\bf 47} (2015) 
art.10, ?rXiv:1408.5344 [gr-qc]
%
\bibitem{1505.07680}
E. Guendelman, E. Nissimov, S. Pacheva, {\it Metric-Independent Volume-Forms in Gravity and Cosmology}, Invited talk at the Memorial "Matey Mateev Symposium", April 2015, Bulgarian Journal of 
Physics, {\bf 42} (2015),  	arXiv:1505.07680 [gr-qc];
%
\bibitem{1507.08878}E. Guendelman, R. Herrera, P. Labrana, E. Nissimov, S. Pacheva, {\it Stable Emergent Universe -- A Creation without Big-Bang}, Astronomische Nachrichten {\bf 336} (2015) 810-814, 
arXiv:1507.08878 [hep-th]
%
\bibitem{1508.02008}E. Guendelman, E. Nissimov, S. Pacheva, {\it Dark Energy and Dark Matter From Hidden Symmetry of Gravity Model with a Non-Riemannian Volume Form}, European Physics Journal C {\bf 
75} (2015) 472-479, arXiv:1508.02008 [gr-qc]
%
\bibitem{1511.07071} E.I. Guendelman, E. Nissimov and S. Pacheva, {\it Unified dark energy and dust dark matter dual to quadratic purely kinetic k-essence}, Eur.Phys.J. C {\bf 76}:90 (2016), 
arXiv:1511.07071[gr-qc] 
%
\bibitem{1603.06231} E. Guendelman, E. Nissimov, S. Pacheva, {\it Gravity-Assisted Emergent Higgs Mechanism in the Post-Inflationary Epoch}, International Journal of Modern Physics D {\bf 
25} 
(2016) 1644008,  	arXiv:1603.06231 [hep-th]
%
\bibitem{1609.06915} Eduardo Guendelman, Emil Nissimov, Svetlana Pacheva.,{\it Quintessential Inflation, Unified Dark Energy and Dark Matter, and Higgs Mechanism},  Bulgarian Journal of Physics 44 
(2017) 15-30, arXiv:1609.06915[gr-qc] 	
%
\bibitem{num_dark} D. Staicova, M. Stoilov, {\em Cosmological Aspects Of A Unified Dark Energy And Dust Dark Matter Model}, Mod. Phys. Lett. A, Vol. 32, No. 1 (2017) 1750006, arXiv:1610.08368 
%
\bibitem{sn}Suzuki et al. (The Supernova Cosmology Project),{\it The Hubble Space Telescope Cluster Supernova Survey: V. Improving the Dark Energy Constraints Above z>1 and Building an 
Early-Type-Hosted Supernova Sample}, ApJ 746, 85 (2012)

\vfill

\end{thebibliography}
\end{document}